\documentclass{emulateapj}
\usepackage{amssymb, amsmath, amsbsy, epsfig, epsf, slashbox}
\usepackage{graphicx}
\usepackage{wrapfig}
\usepackage{color}

\usepackage[T1]{fontenc}
\usepackage{float}
\usepackage{amsthm}
\usepackage{algorithm}
\usepackage{algorithmic}
\usepackage{epstopdf}
\usepackage{hyperref}

\newcommand{\ha}{{\rm H\ensuremath{\alpha}}}
\newcommand{\hb}{{\rm H\ensuremath{\beta}}}

\newcommand{\oiiil}{[O\,{\footnotesize III}] $\lambda$5007}

\newcommand{\niir}{[N\,{\footnotesize II}] $\lambda$6583}

\newcommand{\siil}{[S\,{\footnotesize II}] $\lambda \lambda$6717, 6731}

\newcommand{\oiii}{[O\,{\footnotesize III}]}

\newcommand{\oii}{[O\,{\footnotesize II}]}

\newcommand{\neiii}{[Ne\,{\footnotesize III}]}
\newcommand{\oi}{{\rm [O\,{\footnotesize I}]}}
\newcommand{\nii}{[N\,{\footnotesize II}]}
\newcommand{\sii}{[S\,{\footnotesize II}]}

\def\lax{{$\mathrel{\hbox{\rlap{\hbox{\lower4pt\hbox{$\sim$}}}\hbox{$<$}}}$}}
\def\gax{{$\mathrel{\hbox{\rlap{\hbox{\lower4pt\hbox{$\sim$}}}\hbox{$>$}}}$}}

\begin{document}
%\begin{CJK}{GBK}{song}

%\slugcomment{Submitted to {\it The Astrophysical Journal Letters}}

\title{ Machine Learning Classifiers for Intermediate Redshift Emission Line Galaxies}

\author{
Kai~Zhang\altaffilmark{1}, David~J.~Schlegel\altaffilmark{1},  Brett H. Andrews\altaffilmark{2}, Johan Comparat\altaffilmark{3,4,5}, Christoph Sch{\"a}fer\altaffilmark{6},Jose Antonio Vazquez Mata\altaffilmark{7}, Jean-Paul Kneib\altaffilmark{6,8}, Renbin Yan\altaffilmark{9}}

\altaffiltext{1}{Lawrence Berkeley National Laboratory, 1 Cyclotron Road, Berkeley, CA 94720, USA}
\altaffiltext{2}{PITT PACC, Department of Physics and Astronomy, University of Pittsburgh, Pittsburgh, PA 15260, USA}
\altaffiltext{3}{Instituto de F{\'i}sica Te{\'o}rica UAM/CSIC, 28049 Madrid, Spain}
\altaffiltext{4}{Departamento de F{\'i}sica Te{\'o}rica, Universidad Aut{\'o}noma de Madrid, 28049 Madrid, Spain}
\altaffiltext{5}{Max-Planck-Institut f{\"u}r extraterrestrische Physik (MPE), Giessenbachstrasse 1, D-85748 Garching bei M{\"u}nchen, Germany}
\altaffiltext{6}{Institute of Physics, Laboratory of Astrophysics, Ecole Polytechnique F{\'e}d{\'e}rale de Lausanne (EPFL), Observatoire de Sauverny, 1290 Versoix, Switzerland}
\altaffiltext{7}{Instituto de Astronom\'ia, Universidad Nacional Aut{\'o}noma de M{\'e}xico, A.P. 70-264, 04510, Mexico, D.F., M{\'e}xico}
\altaffiltext{8}{Aix Marseille Universit{\'e}, CNRS, LAM (Laboratoire d'Astrophysique de Marseille) UMR 7326, 13388, Marseille, France}
\altaffiltext{9}{Department of Physics and Astronomy, University of Kentucky, 505 Rose Street, Lexington, KY 40506, USA}

\email{zkdtckk@gmail.com}
\shorttitle{ML Classifiers for Intermediate--z ELGs}
\shortauthors{Zhang et al.}

\begin{abstract}

Classification of intermediate redshift ($z$ = 0.3--0.8) emission line galaxies as star-forming galaxies, composite galaxies, active galactic nuclei (AGN), or low-ionization nuclear emission regions (LINERs) using optical spectra alone was impossible because the lines used for standard optical diagnostic diagrams:  \nii, \ha, and \sii\ are redshifted out of the observed wavelength range.
In this work, we address this problem using four supervised machine learning classification algorithms: $k$-nearest neighbors (KNN), support vector classifier (SVC), random forest (RF), and a multi-layer perceptron (MLP) neural network.
For input features, we use properties that can be measured from optical galaxy spectra out to $z < 0.8$---\oiii/\hb, \oii/\hb, \oiii\ line width, and stellar velocity dispersion---and four colors ($u-g$, $g-r$, $r-i$, and $i-z$) corrected to $z=0.1$.
The labels for the low redshift emission line galaxy training set are determined using standard optical diagnostic diagrams.
RF has the best area under curve (AUC) score for classifying all four galaxy types, meaning highest distinguishing power.
Both the AUC scores and accuracies of the other algorithms are ordered as MLP$>$SVC$>$KNN.
The classification accuracies with all eight features (and the four spectroscopically-determined features only) are 93.4\% (92.3\%) for star-forming galaxies, 69.4\% (63.7\%) for composite galaxies, 71.8\% (67.3\%) for AGNs, and 65.7\% (60.8\%) for LINERs.
The stacked spectrum of galaxies of the same type as determined by optical diagnostic diagrams at low redshift and RF at intermediate redshift are broadly consistent.
Our publicly available code\footnote{\url{https://github.com/zkdtc/MLC_ELGs}} and trained models will be instrumental for classifying emission line galaxies in upcoming wide-field spectroscopic surveys.

\end{abstract}


\section{Introduction}
\label{intro.sec}
Accurate classification of emission line galaxies is critical because the different types of emission line galaxies correspond to different underlying excitation and ionization conditions.
Applying an analysis technique intended for one type of galaxy on another type can produce qualitatively incorrect results (e.g., applying a metallicity calibration on an AGN) because of the built in assumptions about excitation and ionization conditions.

Standard optical diagnostic diagrams, such as the BPT (Baldwin, Philips, \& Terlevich 1981) or VO87 (Veilleux \& Osterbrock 1987) diagrams, are widely used to classify low redshift emission line galaxies into star-forming galaxies, composite galaxies, AGNs, and LINERs.
These diagnostic diagrams use the \oiii/\hb, \nii/\ha, \sii/\ha, and/or \oi/\ha\ lines ratios and some demarcation criteria (e.g., Kauffmann et al. 2003; Kewley et al. 2006).
The advent of large optical spectroscopic surveys like the Sloan Digital Sky Survey (SDSS; York et al. 2000), 2dF (Boyle et al. 2000), and LAMOST has enabled the classification of hundreds of thousands of low redshift ($z<0.3$) emission line galaxies.

Classifying intermediate ($z>0.3$) emission line galaxies is significantly more difficult because the optical spectral features used in the BPT diagram are not captured in optical spectra at these redshifts.
Obtaining the rest-frame optical spectra to apply the BPT diagram requires getting rare and expensive infrared spectra (Trump et al. 2013; Kewley et al. 2013a,b; Azadi et al. 2017).
Classifying intermediate redshift galaxies using only optical spectral and photometric information will enable a wide range of emission line galaxy science with upcoming Stage-IV optical spectroscopic surveys like Dark Energy Spectroscopic Instrument (DESI, Levi et al. 2013), Subaru Prime Focus Spectrograph (PFS; Takada et al. 2014; Tamura et al. 2016), and the 4-metre Multi-Object Spectroscopic Telescope (4MOST; de Jong et al. 2012).

Currently, there are dozens of classification diagrams developed only using parameters available from optical spectra.
Typically, these methods use the fact that AGNs reside exclusively in massive, fast-rotating galaxies and have strong high-ionization lines while star-forming galaxies are less massive, rotate slower, and have lower ionization states.
Some examples of these diagrams include:
\begin{itemize}
    \item the EW(\oii) vs.~EW(\oiii) diagram (Tresse et al. 1996; Rola et al. 1997);
    \item the DEW diagram, which uses 4000 \AA\ break $D_n(4000)$, EW(\oii $\lambda$3727) and EW(\neiii $\lambda$3870) (Stasi\'{n}ska et al. 2006);
    \item diagrams using $g-z$, \neiii, and \oii\ (Trouille et al. 2011);
    \item the $H$-band absolute magnitude vs.~\oiii/\hb\ diagram (Weiner et al. 2006);
    \item the \oii/\hb\ vs.~\oiii/\hb\ diagram (Lamareille 2010);
    \item the $U-B$ color vs.~\oiii/\hb\ diagram (Yan et al. 2011);
    \item the mass--excitation diagnostic (MEx), which uses the stellar mass vs.~\oiii/\hb\ diagram (Juneau et al. 2011, 2013);
    \item the $D_n(4000)$ vs.~\oiii/\hb\ diagram (Marocco et al. 2011); and
    \item the kinematic--excitation diagram (KEx), which uses \oiii\ line width vs.~\oiii/\hb\ (Zhang \& Hao 2018).
\end{itemize}
These diagnostic diagrams generally separate star-forming galaxies and AGNs well, but none of them classify emission line galaxies into the four subtypes that the BPT produces: star-forming galaxies, composite galaxies, AGNs, and LINERs.
Composite galaxies and LINERs are heavily mixed with star-forming galaxies or AGNs on these diagrams.
In this work, we explore the potential for machine learning algorithms to provide accurate 4-class classifications using input features from optical spectra and photometric colors.

In recent years there has been an explosion in the number of applications of machine learning techniques to astronomical problems (see Acquaviva 2019 for a review).
While some studies have used unsupervised algorithms (e.g., Hocking et al. 2018), supervised algorithms have proven to be even more powerful.
The accuracy of neural networks, especially deep convolutional neural networks,  to classify astronomical images has improved dramatically since the early work by de la Calleja \& Fuentes (2004).
For instance, Dieleman et al.~(2015) used a deep convolutional neural network for classifying galaxies using human-labeled images from the Galaxy Zoo project (Lintott et al. 2011) that out-performs experts.
Deep neural networks are also well-suited for identifying strong lens systems in galaxy images because these systems can be robustly simulated even though they are rare in nature (Jacobs et al. 2017, 2019a,b; Petrillo et al. 2017, Pourrahmani et al. 2018; Metcalf et al. 2018; Huang et al. 2019).

Despite the excitement surrounding deep convolutional neural networks, classical supervised machine learning algorithms are often more accurate for problems with relatively few input features, such as classifying emission line galaxies from optical spectral features and photometric colors.
In this paper, we use several such algorithms: K-nearest neighbors (KNN), support vector classifier (SVC), random forest (RF), and a multi-layer perceptron neural network (MLP-NN).

The layout of the paper is as follows.
Section 2 describes the selection and labeling of training, test, and target samples.
Section 3 discusses the selection of input features.
Section 4 compares the performance of our four supervised learning algorithms for classifying low redshift emission line galaxies.
Section 5 describes the application of the trained models to intermediate redshift galaxies.
Section 6 contains our main conclusions.
We use a cosmology with $H_{\rm 0}$ = 70 km\,s$^{-1}$\,Mpc$^{-1}$, $\Omega_{\rm m}$ = 0.3, and
$\Omega_{\rm \Lambda}$ = 0.7 throughout this paper.

\section{Sample}
\label{sample.sec}
\subsection{z$<$0.32 Training and Test Samples}
We apply the following criteria to the SDSS-IV DR15 (Blanton et al. 2017; Aguado et al. 2019)  Extended Baryon Oscillation Spectroscopic Survey (eBOSS; Dawson et al. 2016) data for selecting the low redshift sample for model training, validation, and testing. The Value Added Catalogue\footnote{\url{https://data.sdss.org/sas/dr14/eboss/spectro/redux/v5_10_0/}}  is used to get all the spectral and photometric data used here. We use the Python implementation\footnote{\url{https://pypi.org/project/kcorrect_python/}} of the kcorrect package (Blanton et al. 2007) to convert the u, g, r, i, z magnitudes to z=0.1 values. In order to get the appropriate SED for a set of galaxy fluxes, kcorrect fits an SED which is a nonnegative linear combination of some small number of carefully chosen templates. The z$<$0.32 emission line galaxies sample is selected according to the following criteria:

\begin{description}
\item [(1)]  0$<$z$<$0.32
\item [(2)]  CLASS='GALAXY'
\item [(3)]  SN(\oii)$>$3, SN(\hb)$>$3, SN(\oiii)$>$3, SN(\niir)$>$3, SN(\ha)$>$3, SN(\siil)$>$3
\item [(4)] \oiii\ emission line width ($\sigma(\oiii)$)>0, stellar velocity dispersion ($\sigma_*$) >0.
\end{description}
The final sample consists of 28,869 galaxies, which we split 70/30 into a training+validation sample (20,208 galaxies) and a test sample (8,661 galaxies). We use $k$-fold splitting ($k=6$ for this work) to divide the training+validation sample, which means the sample is split into 6 equal sub-samples and each time one subsample is used as validation sample while the remaining 5 are training samples. This method gives an estimation of error introduced by sample variation and reduces this error in the final prediction by averaging over all $k$ models.

\subsection{Data Labels}
\begin{figure*}
\includegraphics[width=1\textwidth]{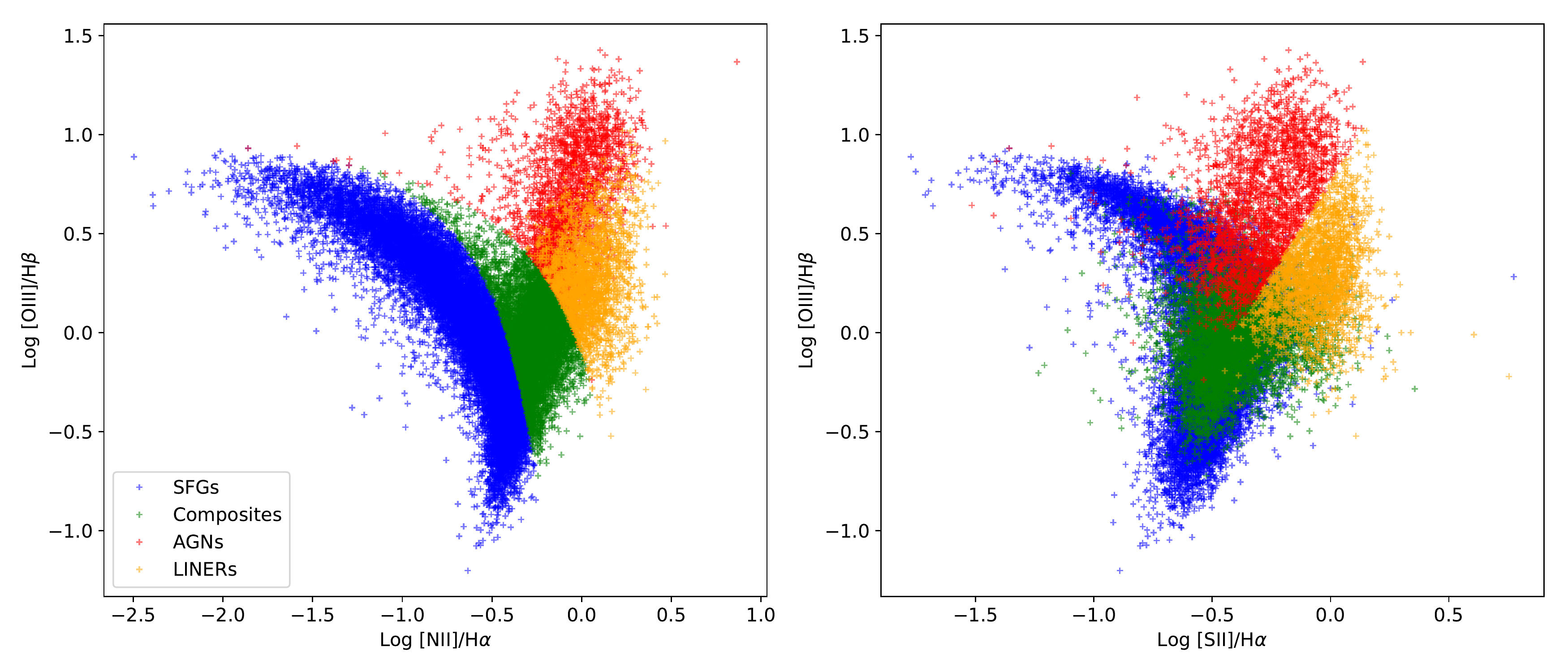}%{bpt.pdf}
\caption{Classification of star-forming galaxies (blue), composites (green), AGNs (red), and LINERs (orange) using the BPT diagram for the z$<$0.32 galaxy sample.  The demarcation lines are from Kauffmann et al. (2003) and Kewley et al. (2006).}
\label{bpt.fig}
\end{figure*}

At z$<$0.32, galaxies can be classified into star-forming galaxies (SFGs), composite galaxies, AGNs, and LINERs using BPT diagrams (Baldwin et al. 1981, Veilleux \& Osterbrock 1987; Kauffmann et al. 2003; Kewley et al. 2006). We use the demarcation lines proposed in Kauffmann et al. (2003) and Kewley et al. (2006) for classification into four subtypes. The distributions of the four subtypes of galaxies in the BPT diagrams are shown in Figure~\ref{bpt.fig}.  Star-forming galaxies, composites, AGNs, and LINERs are denoted in blue, green, red, and orange, respectively. There are 17,073 SFGs, 6,826 composites, 2,566 AGNs, and 2,399 LINERs.

\subsection{0.32$<$z$<$0.8 Emission Line Galaxies Sample}
\label{highz_sample.sec}
Our goal is to classify intermediate redshift (0.32$<$z$<$0.8) ELGs. The intermediate redshift sample is selected based on the following criteria:

\begin{description}
\item [(1)]  0.32$<$z$<$0.8
\item [(2)]  CLASS='GALAXY'
\item [(3)]  SN(\oii)>3, SN(\hb)>3, SN(\oiii)>3
\item [(4)] \oiii\ emission line width ($\sigma(\oiii)$)>0, stellar velocity dispersion ($\sigma_*$) > 0.
\end{description}
The final sample consists of 49,272 galaxies. We use the kcorrect package to convert the u, g, r, i, z magnitudes to z=0.1 values.

\section{Input Features}
\label{features.sec}
In machine learning terminology, features are the input parameters. We use `features' as the standard term here.
We select \oiii/\hb, \oii/\hb, \oiiil\ line width $\sigma_{\oiii}$, stellar velocity dispersion $\sigma_*$, u-g, g-r, r-i, and i-z as the input features for classification. \oiii/\hb, \oii/\hb, \oiiil\ line width $\sigma_{\oiii}$, stellar velocity dispersion $\sigma_*$ can be easily measured from optical spectra of z$<$0.8 galaxies and can be measured out to even higher redshift if NIR spectra are available. The SDSS imaging survey\footnote{\url{https://www.sdss.org/dr12/imaging/}} provides $u-g$, $g-r$, $r-i$, and $i-z$ colors for 14,055 square degrees of the sky. If a source is not detected, we  use its magnitude upper limit because an upper limit is still informative. The $g$, $r$, and $z$ photometry is supplemented using The Legacy Surveys \footnote{\url{http://legacysurvey.org}} Data Release 7 (Dey et al. 2019) values if available. The Legacy Surveys are producing an inference model catalog of the sky from a set of optical and infrared imaging data, comprising 14,000 deg$^2$ of the extragalactic sky visible from the northern hemisphere in three optical bands (g, r, and z) and four infrared bands. These input features are selected from previous works of intermediate redshift emission line galaxies diagnostic diagrams (e.g., Lamareille 2010; Yan et al. 2010; Zhang \& Hao 2018). They are chosen because \oii, \hb, and \oiii\ are the strongest emission lines at rest-frame wavelengths shorter than 5010 \AA.  Stellar velocity dispersion can be well-measured using continuum fitting, and the $u$, $g$, $r$, $i$, and $z$ broad band magnitudes have high signal-to-noise ratios. We do not use stellar mass measurements (Juneau et al. 2010) because these are  derived values with typical errors of 0.3$-$0.4~dex, and $\sigma_*$  and $\sigma_{\oiii}$ already contain information about the mass of a galaxy. We chose not to use $D_n(4000)$ because it is less informative than the u-g color. The \neiii\ line is not selected because of its weakness. One could add more input features, like colors using other bands, more emission lines ratios, or equivalent widths, and this might or might not improve the classification accuracy. For this paper, we just use the 8 input features to set a baseline.

\begin{figure*}
\includegraphics[width=1.0\textwidth]{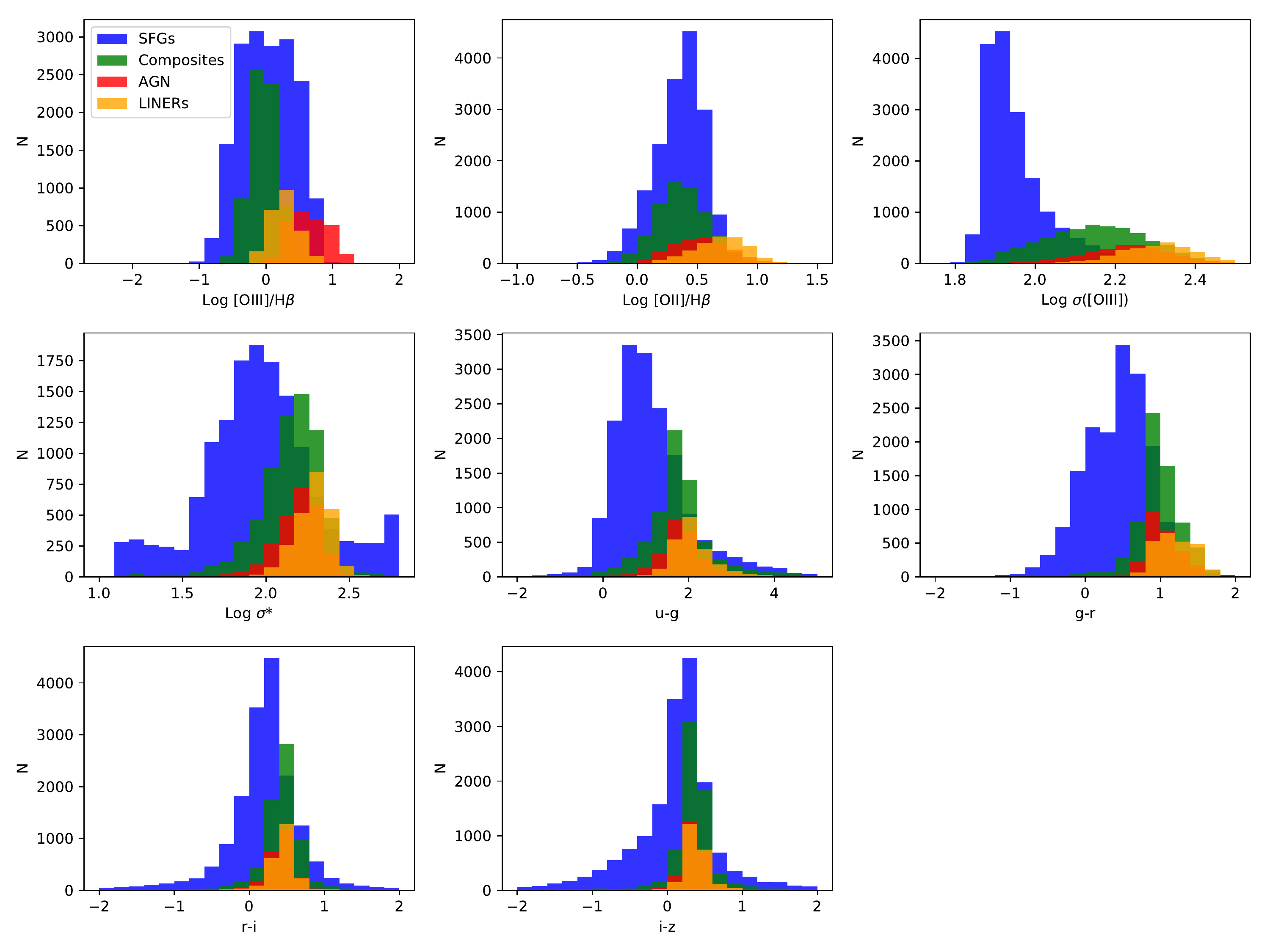}%{histgram.pdf}%feature.pdf
\caption{The distribution of the 8 input features for the four subtypes of ELGs for the whole z$<$0.32 sample.}
\label{input1.fig}
\end{figure*}

\begin{figure}
\includegraphics[width=0.5\textwidth]{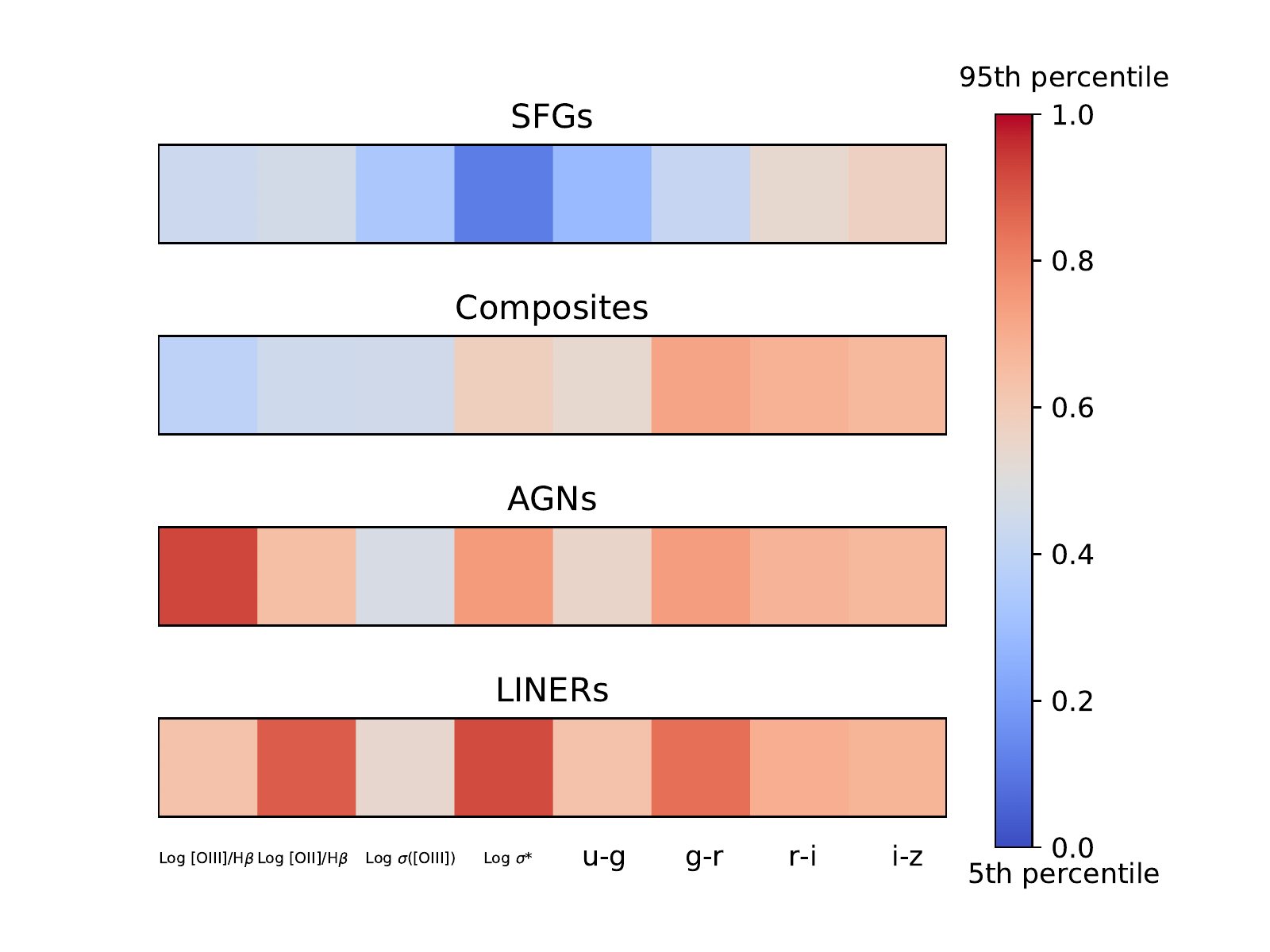}%feature.pdf
\caption{The median values of the 8 input features for the four subtypes of ELGs for the whole z$<$0.32 sample to illustrate the distinguishing power of each feature. All features are normalized to the 5--95 percentile range. The median values of the four subtypes are easily distinguished from each other using the 8 features here.  }
\label{input2.fig}
\end{figure}

In Figure~\ref{input1.fig}, we show the distribution of the 8 features for the four subtypes for the low redshift galaxy sample. Figure~\ref{input2.fig} shows the median values of the 8 input features for the four subtypes ELGs for the whole z$<$0.32 sample to illustrate the distinguishing power of each feature. All features are normalized to the 5--95 percentile range. SFGs are characterized by low \oiii/\hb, \oii/\hb, $\sigma_{\oiii}$, $\sigma_{*}$, u-g, g-r, and r-i. Thus, they are clustered in a very small volume in the 8 dimensional parameter space. AGNs are characterized by extremely high \oiii/\hb, and all other 7 features are near the median. LINERs show the highest \oii/\hb, $\sigma_{*}$, and g-r color. Composites have median values for all 8 features between 0.4 and 0.6. On average, the four subtypes are easily distinguished using the 8 features. However, we do not consider the dispersion of each feature, so the four subtypes could still be heavily mixed with each other in parameter space and thus not 100\% separable, as shown later in the paper.

\section{Model Training and Performance}
\label{result.sec}

We use several popular supervised learning methods and quantify their classification accuracy. For each method, we briefly introduce the algorithm, fine-tune the hyperparameters, and report its performance. We note that performance on our data set is not necessarily indicative of performance on other data sets because these methods are sensitive to the particulars of the data set. The discussion is strictly confined to the data we use here and the models we use.

The four subtypes of emission line galaxies are not equally represented in the final sample. There are 17,073 SFGs, 6,826 composites, 2,566 AGNs, and 2,399 LINERs. If we directly feed the imbalanced training sample into a model, it will be biased in favor of the over-represented subtypes and biased against the under-represented subtypes. In our case, the model would excel at selecting SFGs but struggle with distinguishing the other three subtypes. To mitigate this problem, we created a new sample equally-weighted across subtypes by randomly selecting galaxies from each subtype.  This is equivalent to giving higher weights to subtypes with fewer instances.

To quantify the performance of each method, the trained classifier is applied to the test sample, and the fraction of correct classification is the accuracy for a specific subtype.
%The mean value of the 4 accuracy is denoted as 'overall accuracy', which is used as the dominant success metric to evaluate the performance of a model.
The receiver operating characteristic (ROC) curve (Metz, 1978; Fawcett, 2006) and area under the ROC curve (AUC) score (Bradley, 1997) are used to quantify the distinguishing power of different classifiers. The confusion matrix of a classifier include:
\begin{description}
\item [(1)]  True Positive (TP)---correct identification.
\item [(2)]  True Negative (TN)---correct rejection.
\item [(3)]  False Positive (FP)---incorrect identification, also called a false alarm or Type I error.
\item [(4)] False Negative (FN)---incorrect rejection, also called a Type II error.
\end{description}
The ROC curve uses the true positive rate (TPR=$\frac{TP}{TP+FN}$) as the y-axis and the false positive rate (FPR=$\frac{FP}{FP+TN}$) as the x-axis. It reflects the tradeoff between TPR and FPR for different thresholds and thus different demarcation hyperplanes. In binary classification, setting the threshold too high will produce a high TPR and high FPR, meaning that a large fraction of true positives get selected but a large number of false positive will be misclassified as well. Setting the threshold too low results in a low TPR and a low FPR, meaning that many false positives are successfully rejected at the cost of not selecting many true positives. A perfect classifier has a TPR=1 and an FPR=0, which produces an AUC score of 1. The worst possible classifier, on the other hand, has a TPR=0 and an FPR=1, which produces an AUC score of 0. Consequently, AUC score is commonly used to evaluate the classification power of a model with higher AUC scores being better. As such, this tool has become standard in optimization scenarios, but applying it to multi-class cases is more challenging. The general idea is to convert the multi-class problem into several binary classification problems using the one-vs.-rest method (Mossman 1999; Srinivasan, 1999; Hand and Till, 2001; Ferri et al., 2009). For each ML technique we describe in the following sections, we present the ROC curves and AUC scores for each galaxy subtype relative to the other three subtypes.

\subsection{k-Nearest Neighbor Method}
\label{knn.sec}
The classification problem presented in this paper is relatively simple and well-suited for classical machine learning methods that have widely available implementations:  $k$-Nearest Neighbors, linear SVC, non-linear SVC, etc. We use $k$-nearest neighbors (KNN) to establish a baseline of classification accuracy because is the most straightforward method to use to make a classification. The classification of a source is determined by the voting results of the $k$ neighbors who are the nearest to the input in the multi-dimensional parameter space.

\subsubsection{KNN Performance}
\begin{figure}
\includegraphics[width=0.5\textwidth]{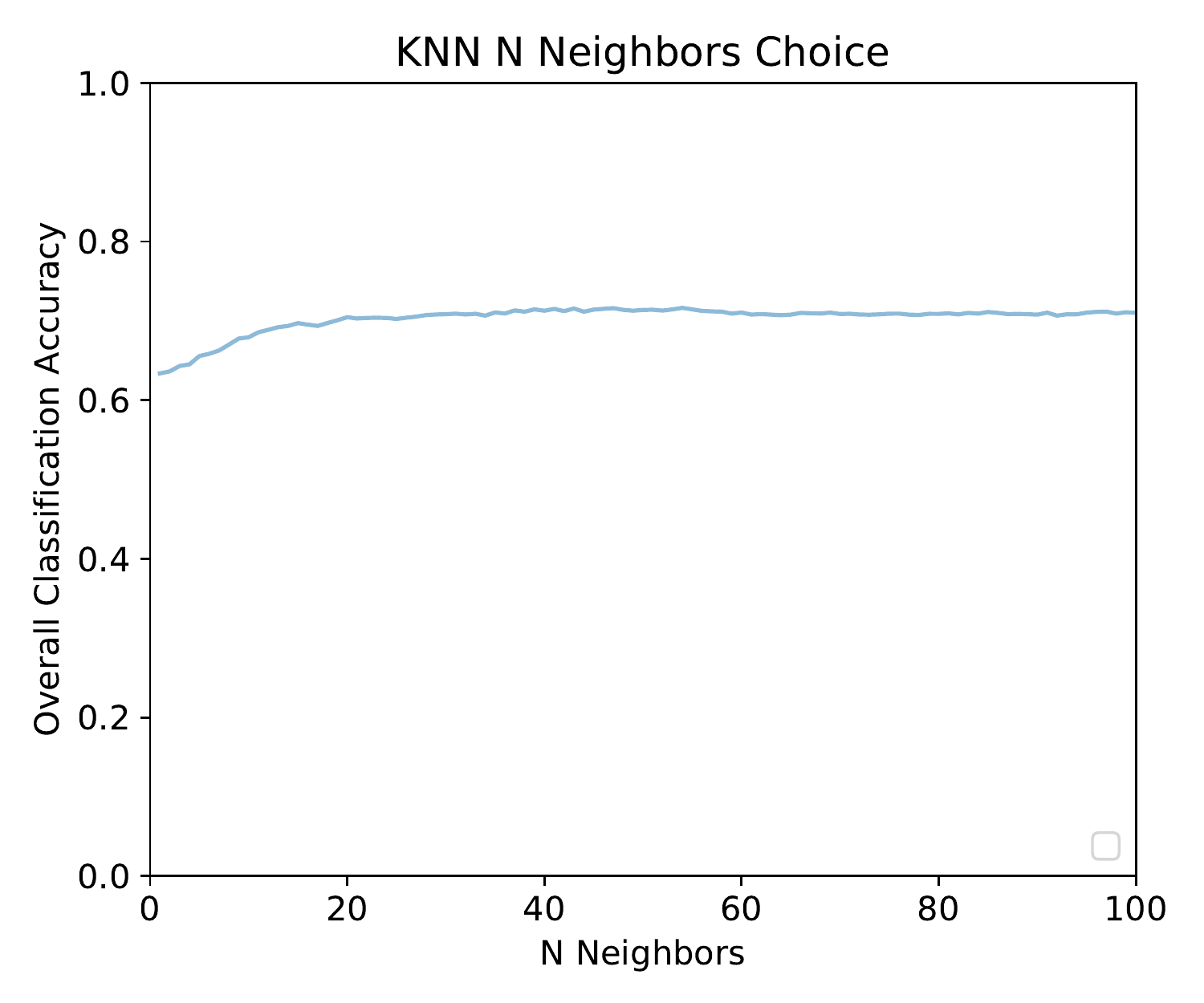}%knn_choose.pdf
\caption{Overall accuracy on the validation sample as a function of the number of neighbors for classification voting ($k$). The best performance is achieved with $k$=54. Therefore, $k$ is set to 54.}
\label{knn_choose.fig}
\end{figure}

\begin{figure*}
\includegraphics[width=1\textwidth]{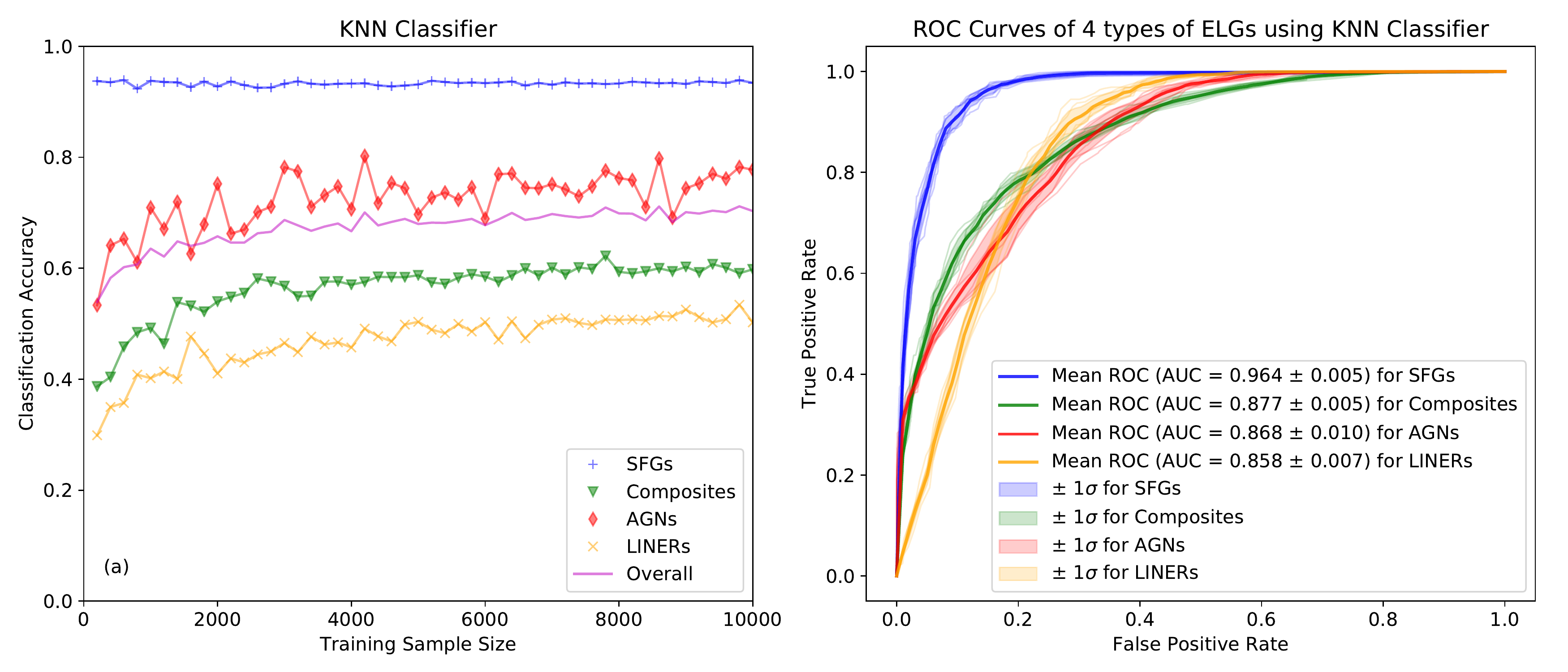} %knn_performance
\caption{Panel (a): The KNN classification accuracy as a function of training sample size for the four subtypes of emission line galaxies. Blue crosses, green triangles, red diamonds and orange x's denote SFGs, composites, AGNs, and LINERs. The magenta line shows the average classification accuracy of the 4 subtypes.  Panel (b): ROC curves and AUC scores for each subtype of ELGs using the $k$-nearest neighbors method.}
\label{knn_performance.fig}
\end{figure*}

We use the KNN implementation in scikit-learn v0.21.2 (Pedregosa et al 2011). The number of neighbors for voting, $k$, is a free hyperparameter. In Figure~\ref{knn_choose.fig}, we plot the overall accuracy as a function of $k$ for the validation sample to determine the optimal $k$ value, which we find to be $k$=54. Thereafter, we trained KNN classifiers using $k$=54.  In panel (a) of Figure~\ref{knn_performance.fig}, we plot the classification accuracy for the four subtypes and the overall accuracy as a function of training sample size. The overall accuracy climbs from 0.5 to about 0.65 when the training sample size reaches 3000 and plateaus after that. The corresponding subtype accuracies are 93.4\%, 59.6\%, 76.0\%, and 51.6\% for SFGs, composites, AGNs, and LINERs. The ROC curves and AUC score for each subtype of ELG using KNN is shown in Panel (b). We use 6-fold cross validation to evaluate the performance.  ``6-fold cross validation'' means we split the training sample into 6 equal subsamples, and each time one subsample is used as the validation sample and the other 5 subsamples are used as training samples. The thin lines are the ROC curves for individual validation subsamples, and the thick lines are the mean ROC curves. The shaded areas are the 1$\sigma$ errors of the mean ROC curves.  KNN is quite good at distinguishing SFGs, composites, AGNs, and LINERs, with AUC scores of 0.964, 0.878, 0.860, and 0.865, respectively. We note that composites, AGNs, and LINERs have similar AUC scores, but they have very different classification accuracies in Figure~\ref{knn_performance.fig}a. This is because the final accuracies are the result of tradeoffs amongst the four subtypes. KNN produces high accuracies for AGNs and composites at the expense of LINER accuracy.

%In panel (b), the completeness of 4 subtypes as a function of training sample size are given. The final completeness are 81\%, 66\%, 34\% and 79\% for SFGs, Composites, AGNs and LINERs.  The classification accuracy shows no correlations with completeness, as shown in panel (c). This is reasonable considering the classification algorithm. The accuracy and completeness are determined by the types of neighbors around a source, which is very stable when training sample size is large enough.  There are no tradeoff between accuracy and completeness here. The accuracies for 4 types show no correlation with each other, as shown in panel (d). The lack of correlations between accuracy and completeness and between accuracies of different subtypes illustrate the KNN algorithm is not separating galaxies based on a demarcation hyperplane, but depend on local neighbors.

\subsection{Support Vector Classifier}
\label{svc.sec}
Support vector machines have been one of the best tools for regression and classification. A support vector classifier (SVC) finds the demarcation plane by maximizing the distance between the hyperplane and the nearest point. Linear SVC assumes that the demarcation hyperplane is linear while non-linear SVC does not make such an assumption. We use the scikit-learn implementation of non-linear SVC for supervised learning on our sample. We also tried the linear SVC method, but it shows significantly lower accuracy for all four subtypes, thus we do not present it in this paper. The lower accuracy for linear SVC could be caused by the non-linear demarcation lines in the 2--3 parameter BPT diagrams, which also translates to our input features.

\subsubsection{SVC Performance}
\begin{figure*}
\includegraphics[width=1\textwidth]{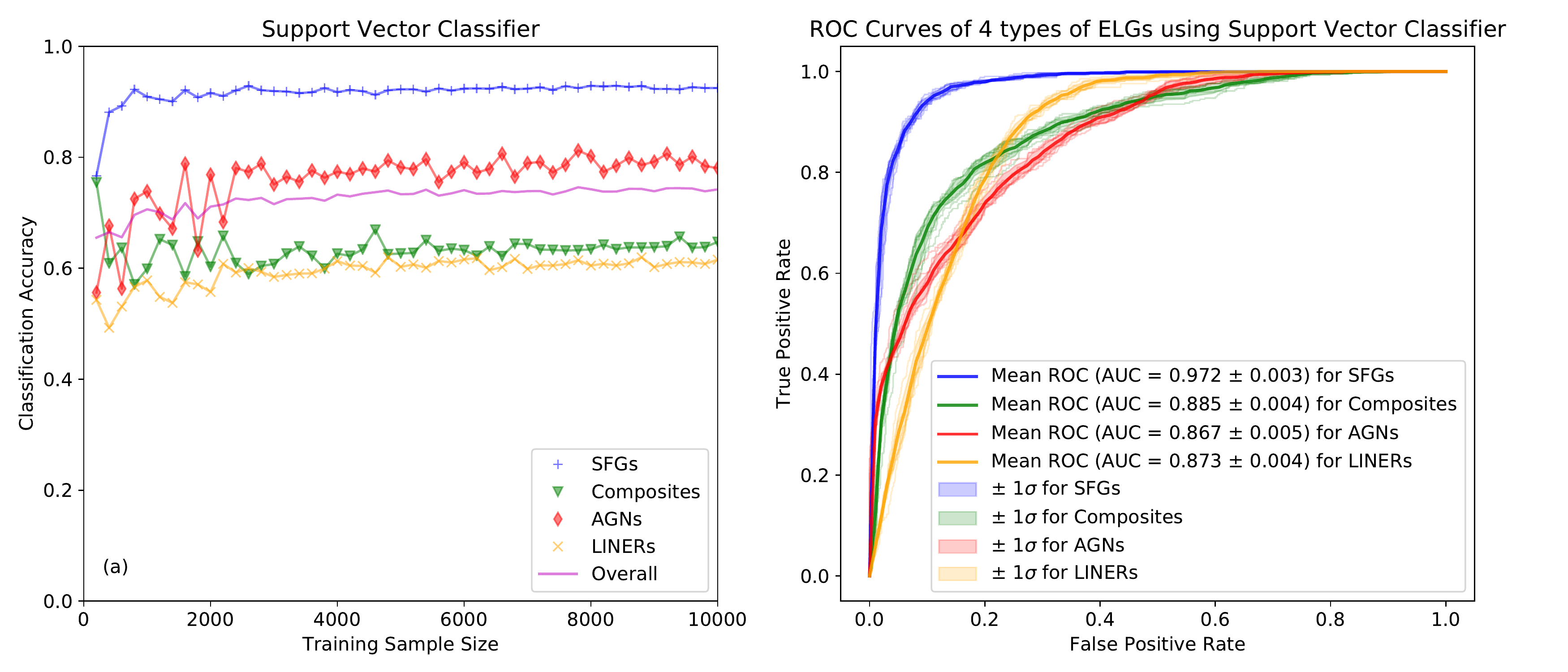} %svc_performance.pdf
\caption{Panel (a): The SVC classification accuracy as a function of training sample size for the four subtypes of emission line galaxies. The legends are the same as Figure~\ref{knn_performance.fig}. The accuracy curves for the 4 subtypes are stable after the training sample size reach 1,000. Panel (b): ROC curves and AUC score for each type of ELGs using SVC.}
\label{svc_performance.fig}
\end{figure*}

The classification accuracy as a function of training sample size is shown in Figure~\ref{svc_performance.fig}. The accuracy curves for the 4 subtypes stabilize after the training sample size reaches 1,000 galaxies (250 per subtype). The final accuracies for SFGs, composites, AGNs, and LINERs are 92.6\%, 63.8\%, 79.4\%, and 60.8\%. The accuracy curves saturate after 1,000 training sources because the separation hyperplane is optimized and more sources do not change the hyperplane significantly.  With a larger training sample, the accuracy hardly changes, thus the accuracy variation is small.
The ROC curves and AUC score for SFGs, composites, AGNs, and LINERs using SVC is shown in Panel (b) of Figure~\ref{svc_performance.fig}. The AUC scores are 0.968 (SFGs), 0.881 (composites), 0.861 (AGNs), and 0.869 (LINERs). Despite similar AUC scores, SVC is much better than KNN for classification accuracy.

\subsection{Random Forest}
\label{rf.sec}
Another popular method for classification is decision trees. A decision tree classifies an object according to a series of criteria. However, a single tree usually introduces a cut in parameter space that is not ideal. The criteria from a single decision tree may not be ideal, so using many decision trees and letting them vote for the classification result produces much better outcomes than a single decision tree. One popular ensemble method is random forest (RF), which creates trees each with a random subset of input features and a random sample of data with replacement. We use the scikit-learn RF implementation with n\_estimators=1000, oob\_score=True, and n\_jobs=-1.

%One of the cutting edge ensemble learning method is gradient boost method. Gradient Boost builds an additive model in a forward stage-wise fashion; it allows for the optimization of arbitrary differentiable loss functions. We use the XBboost (Chen \& Guestrin 2016) implementation here. The hyperparameter setup is: max\_depth=6, n\_estimators=300, reg\_lambda=1, learning\_rate=0.1

\subsubsection{Random Forest Performance}
\begin{figure*}
\includegraphics[width=1\textwidth]{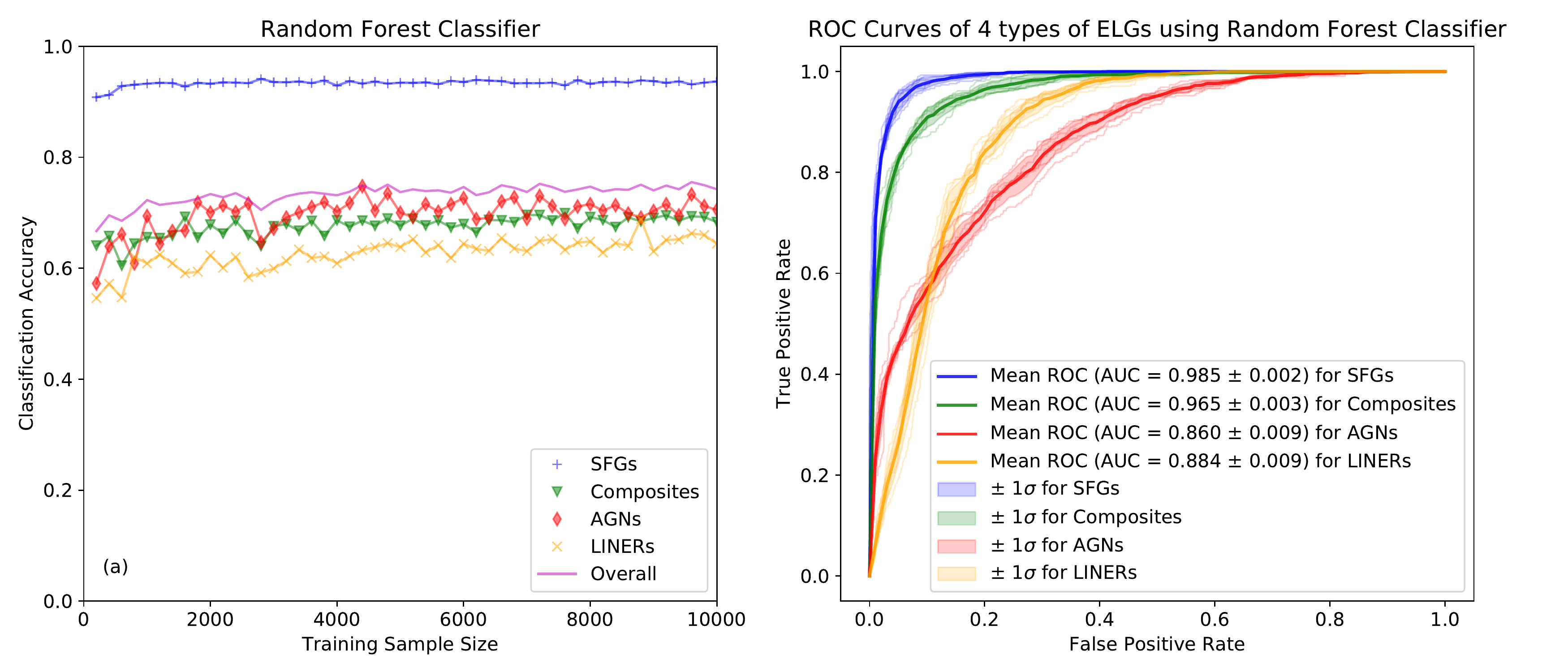} %rf_performance.pdf
\caption{Panel (a): The Random Forest classifier accuracy as a function of training sample size for the 4 subtypes of emission line galaxies. The legends are the same as Figure~\ref{knn_performance.fig}. The accuracy curves for the 4 subtypes are stable after the training sample size reaches 10,000 galaxies. Panel (b): ROC curves and AUC scores for each subtype of ELG using the Random Forest method.}
\label{rf_performance.fig}
\end{figure*}

The classification accuracy as a function of training sample size is shown in Figure~\ref{rf_performance.fig}. With a very small sample size of about only 100 sources, the random forest classifier gives a good overall accuracy of $\sim$0.65. The accuracy keeps climbing with increasing sample size and stabilizes at 0.75 at 10,000 training sources. The final accuracy for SFGs, composites, AGNs, and LINERs are 93.4\%, 69.4\%, 71.8\%, and 65.7\%, respectively. The AUC scores for SFGs, composites, AGNs, LINERs are 0.985, 0.966, 0.876, 0.897, respectively---a big improvement over the KNN and SVC methods.

%The corresponding completenesses are 90\%, 72\%, 42\% and 61\%. The classification accuracy shows no correlations with completeness except for composites. Composites show a positive correlation between classification accuracy and completeness. This is very interesting, because we expect a negative correlation if a smooth hyperplane exists to tradeoff accuracy and completeness. The presence of a positive correlation means the as the training sample enlarges, the model learns how to precisely select individual composite based on more decision trees. Principally, this is similar to KNN method in Section~\ref{knn.sec}, but the decision is made through a much more complex logic, thus the precision is significantly higher than KNN.

\subsection{Importance of Individual Input Parameters}
Feature importance measures the distinguishing power of each feature, and opens the possibility of dropping the least important features without significantly sacrificing performance. A benefit of using gradient boosted methods is that it is straightforward to calculate importance scores for each attribute after the boosted trees are constructed. For a single decision tree, the importance is calculated by multiplying the amount that each attribute split point improves the performance measure by the number of observations for which the node is responsible. The final feature importances are the average values over all decision trees within the model. Generally, the more an attribute is used to make key decisions with decision trees, the higher its relative importance.\footnote{For details on how importance in decision tree is calculated see Section 10.13.1 ``Relative Importance of Predictor Variables'' of the book ``The Elements of Statistical Learning: Data Mining, Inference, and Prediction,'' page 367.}

\begin{figure}
\includegraphics[width=0.5\textwidth]{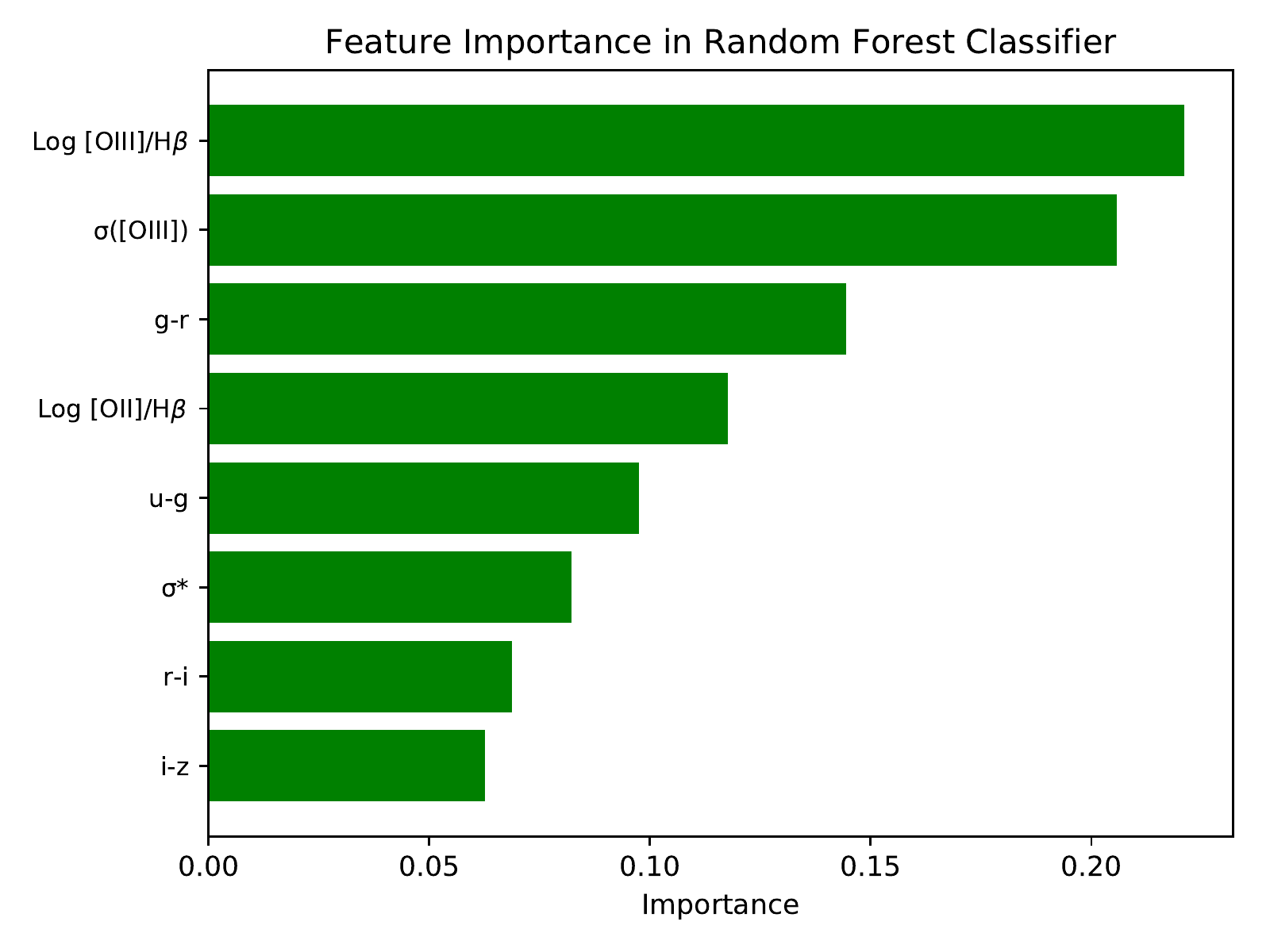} %rf_importance.pdf
\caption{The importances of the 8 features for the random forest classifier. The top 3 most important features are \oiii/\hb, $\sigma_{\oiii}$, and g-r. \oii/\hb, u-g, and $\sigma_{*}$ are ranked 4--6. r-i and i-z are the least important. }
\label{rf_importance.fig}
\end{figure}

We show the importances of the 8 features in Figure~\ref{rf_importance.fig}. The top 3 most important features are \oiii/\hb, $\sigma_{\oiii}$, and g-r. \oii/\hb, u-g, and $\sigma_{*}$ are ranked 4--6.  r-i and i-z are the least important with importance values about 1/3 of the most important feature (\oiii/\hb).  These results are impressive given that the machine learning methods were not provided any physics principles but rather tell us what features are the most important purely from the data.  Calculating feature importance can be extremely helpful for data sets with huge numbers of features whose physical meanings and connections are not well understood.
%The importance analysis here tells us: the most important features to distinguish are high ionization line strength (\oiii/\hb), low-ionization line strength (\oii/\hb), and galaxy kinematics ($\sigma_{\oiii}$. Colors are relatively less important.

\subsection{Neural Network}
\label{mlp.sec}
%CNN has been shown to out perform SVM or other methods on the most complex computer vision classification problems like IMAGENET\footnote{http://www.image-net.org} (Russakovsky et al. 2015).
Finally, we apply a neural network to our classification problem. A neural network is combination of layers of neurons, just like our brain. The parameters of a neuron (weight and bias for a linear neuron) can be adjusted through the learning process. A loss function is defined to quantify how poorly the model is at making prediction. To improve the prediction accuracy of the model, the prediction error is back-propagated through the network, and the model is updated accordingly. The model predictions improve with additional data.

\subsubsection{Neural Network Setup}
Usually, the deeper the network, the better its performance. A two layer convolutional neural network is powerful enough to achieve 98\% accuracy in classifying hand-written digits (0--9) from the MNIST data set. Our emission line galaxy classification problem is less complex than the MNIST task, so we use the multi-layer perceptron (MLP) classifier implementation in the scikit-learn neural\_network package with the L2 penalty parameter alpha set to 0. Other parameters are set to the default values: one hidden layer with 100 neurons and rectified linear unit function (RELU) as the activation function. The learning rate is 0.001, and maximum number of iterations is set to 200. Our model optimizes the log-loss function using stochastic gradient descent method by setting the solver to `adam' (Kingma \& Ba 2015). We set the exponential decay rate for estimates of first and second moment vector in `adam' to 0.9 and 0.999, respectively.

\subsubsection{Multi-layer Perceptron Classifier Performance}
\begin{figure*}
\includegraphics[width=1\textwidth]{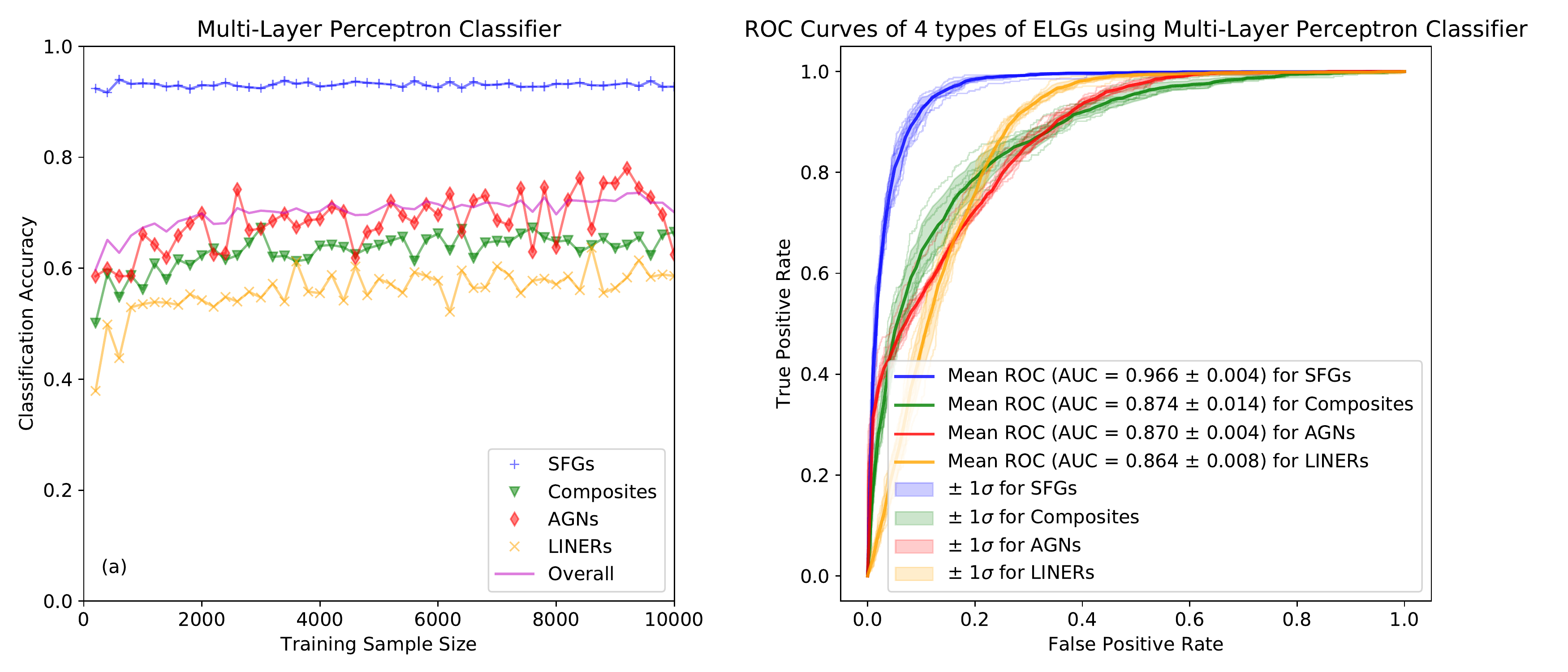} % mlp_performance.pdf
\caption{Panel (a): The MLP classifier classification accuracy as a function of training sample size for the four subtypes of emission line galaxies. Legends are the same as Figure~\ref{knn_performance.fig}. The accuracy curves for the four subtypes are stable after the training sample size reaches 2,000. Panel (b): ROC curves and AUC scores for each subtype of ELGs using the multi-layer perceptron classifier. }
\label{mlp_performance.fig}
\end{figure*}

The classification accuracy as a function of training sample size is shown in Figure~\ref{mlp_performance.fig}.
With increasing sample size, the accuracy keeps climbing and stabilizes at 0.75 at 2,000 training sources. The final accuracy for SFGs, composites, AGNs, and LINERs are 93.7\%, 68.8\%, 76.8\%, and 61.1\%, respectively. The AUC scores for SFGs, composites, AGNs, LINERs are 0.964, 0.874, 0.867, 0.864, respectively, which is very close to the random forest classifier.

\subsection{Performance Comparison}
\begin{figure*}
\includegraphics[width=1\textwidth]{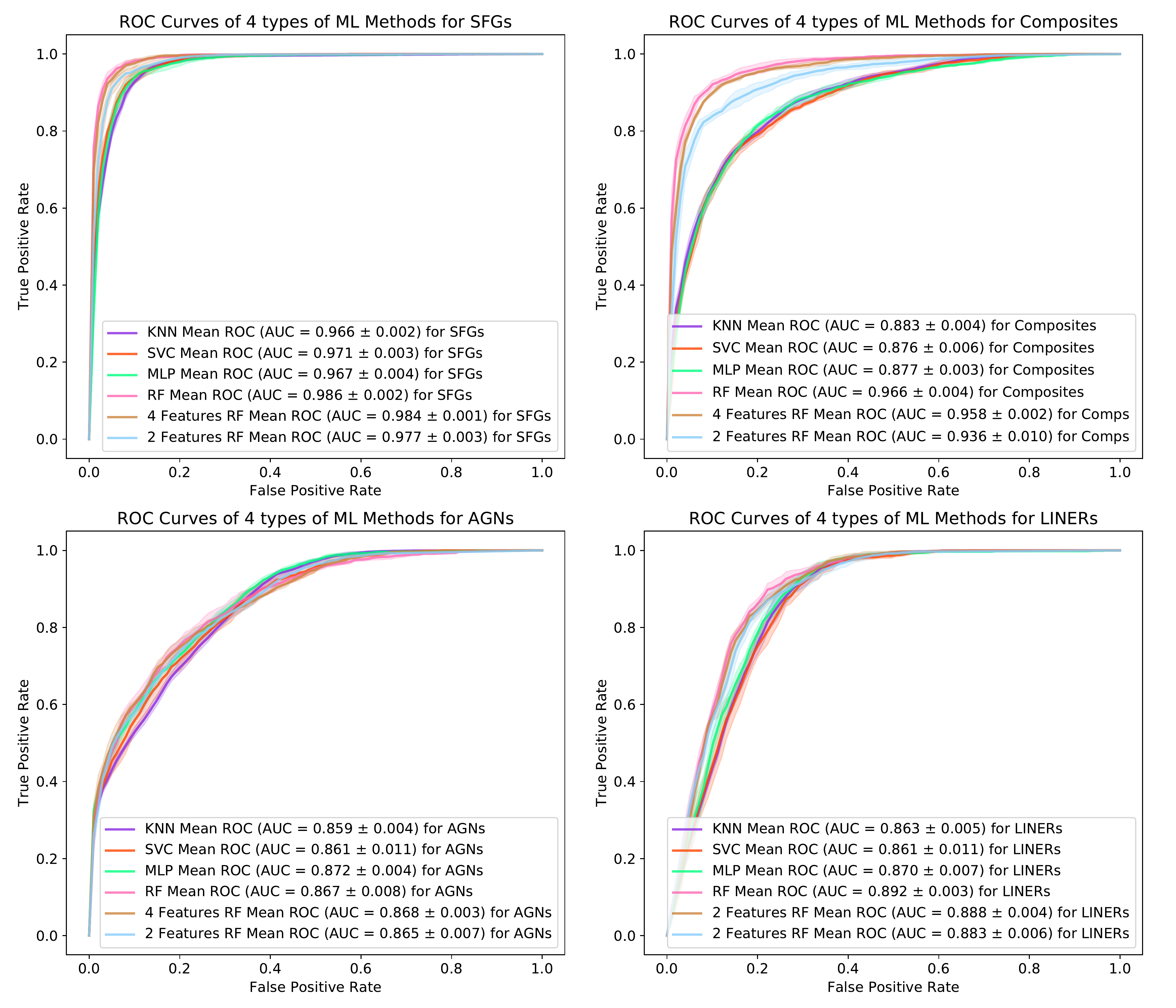}%{roc_plot.pdf}
\caption{The comparison of ROC curves and AUC scores for each subtype of ELGs with the four ML methods. The random forest classifier achieves significantly higher AUC scores for SFGs, Composites and LINERs than the other methods, and it performs similarly to the other methods in classifying AGNs.}
\label{roc_plot.fig}
\end{figure*}

The AUC scores and accuracies for the four subtypes are given in Table~\ref{auc.tab} and Table~\ref{accuracy.tab}. The ROC curves and AUC scores for the four subtypes of ELGs for each ML method is shown in Figure~\ref{roc_plot.fig}. The average AUC score for the four subtypes are 0.892, 0.895, 0.931 and 0.892 for KNN, SVC, RF, and MLP, respectively. The rank in average accuracy is the same as the rank of AUC scores, 70.2\%, 74.1\%, 75.1\%, and 75.0\% for KNN, SVC, RF, and MLP. However, the different methods settle on different demarcation hyperplanes, resulting in different preferences. MLP has the highest accuracy for star-forming galaxies, and SVC has the highest accuracy for AGNs. RF achieves the highest accuracies for composites and LINERs. These differences reflect the different tradeoffs of the four algorithms. For robustness and performance stability, we favor the Random Forest Classifier as the optimal method.  The confusion matrix of the random forest classifier is given in Table~\ref{confusion.tab}. Star-forming galaxies are unlikely to be confused with the other subtypes. Composites have a 23.8\% probability to be confused with star-forming galaxies. AGNs have 18.8\%,  8.9\%, and 1.8\% probabilities to be classified as LINERs, composites, and star-forming galaxies, respectively. LINERs are most likely to be misclassified as composites with 28.4\% probability, and 6.7\% and 0.4\% probabilities to be misclassified as AGNs and star-forming galaxies, respectively.

\subsection{Classification Using Only Spectroscopic Features}
\begin{figure*}
\includegraphics[width=1\textwidth]{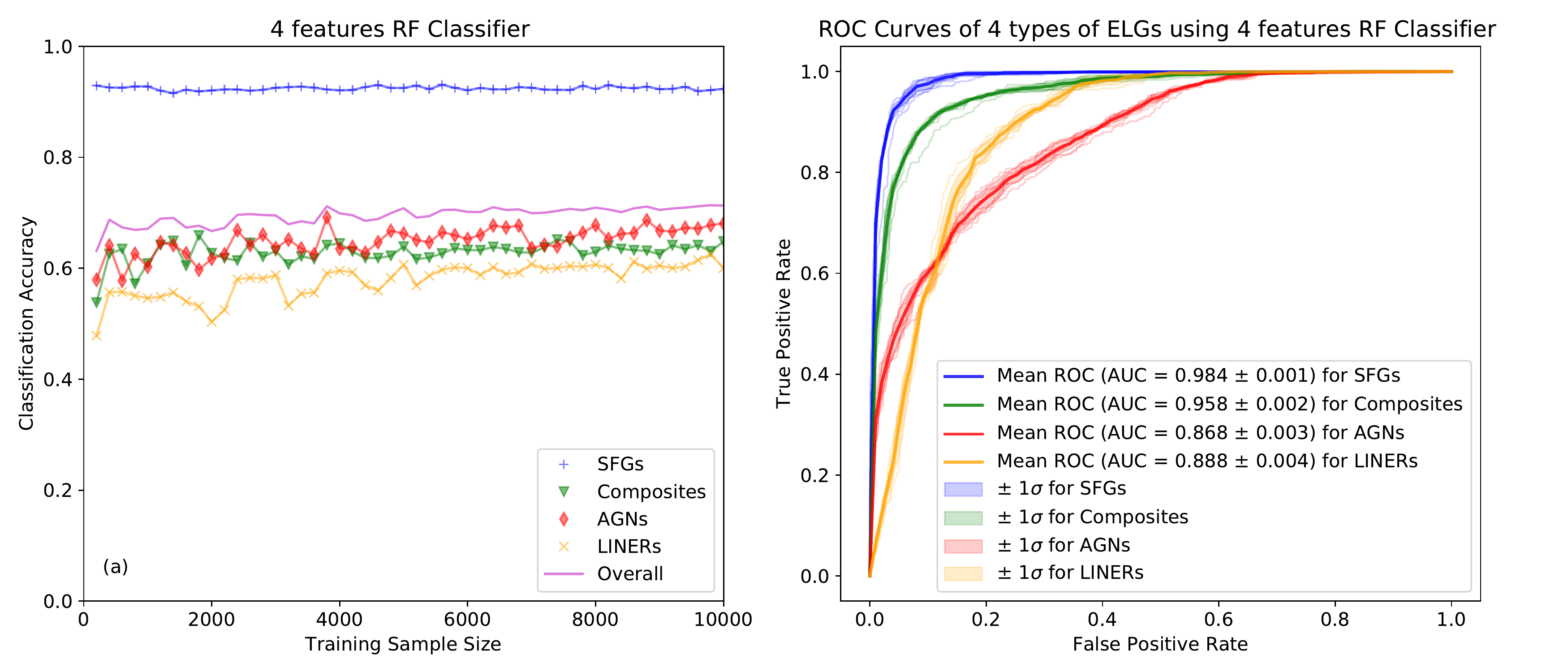}%{rf4_performance.pdf}
\caption{ Panel (a): The random forest classifier accuracy as a function of training sample size for 4 subtypes of emission line galaxies, using only 4 features: \oiii/\hb, \oii/\hb, $\sigma(\oiii)$), and stellar velocity dispersion. Legends are the same as Figure~\ref{knn_performance.fig}. The accuracy curves for 4 subtypes are stable after the training sample size reaches 10,000 galaxies. Panel (b): ROC curves and AUC scores for each type of ELGs using the random forest method and only 4 spectroscopic features.   }
\label{rf_performance_4f.fig}
\end{figure*}

%It is tantalizing to add more features to the model to achieve higher distinguishing power. After all, intuitively more information should produce smarter model. We add z-W1 color to the input features and rerun the training process. However, the performance of the model using all methods drop. For Random Forest Classifier, the AUC scores drop by 0.02 for all subtypes. This is interesting, because z-W1 seem to counter-balance the distinguishing power of some other parameters. We leave the detailed discussion in this aspect for future works.
It is useful to construct a classifier based solely on spectroscopic features because imaging data is not always available. We reduce the feature set to \oiii/\hb, \oii/\hb, $\sigma(\oiii)$), and stellar velocity dispersion ($\sigma_*$), and use the same training sample as in Section \ref{result.sec}.  We use a random forest classifier to see how well it performs with 4 features compared to 8 features. The accuracy curves and ROC curves are given in Figure~\ref{rf_performance_4f.fig}. The AUC scores drop from 0.985, 0.966, 0.876, and 0.897 to 0.981, 0.952, 0.870 and 0.890 for SFGs, composites, AGNs, and LINERs, respectively. Thus, the colors do help in classification, but reducing the feature set to only spectroscopic features does not degrade the classification performance significantly. It would be ideal to have 8 features for the classification ELGs, but using only 4 spectroscopic features \oiii/\hb, \oii/\hb, $\sigma(\oiii)$ and stellar velocity dispersion ($\sigma_*$) can give a very similar result. 

\subsection{Machine Learning Classifications on the BPT, Kinematic--Excitation, and Mass--Excitation diagrams}
\begin{figure*}
\includegraphics[width=1\textwidth]{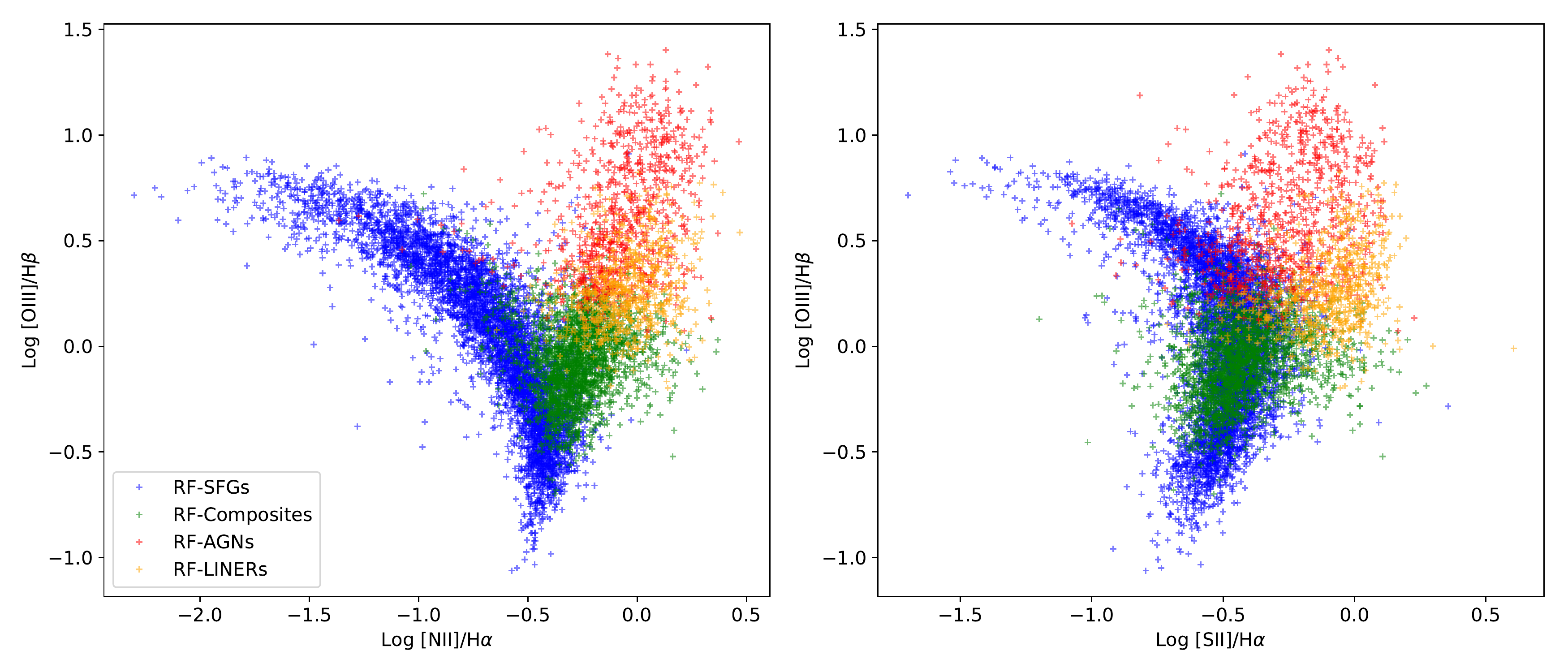}%{new_bpt.pdf}
\caption{ The RF-classified z$<$0.32 galaxies of four subtypes on the BPT diagram. The RF classifier does an excellent job of reproducing the BPT diagram classification.    }
\label{new-bpt.fig}
\end{figure*}

\begin{figure*}
\includegraphics[width=1\textwidth]{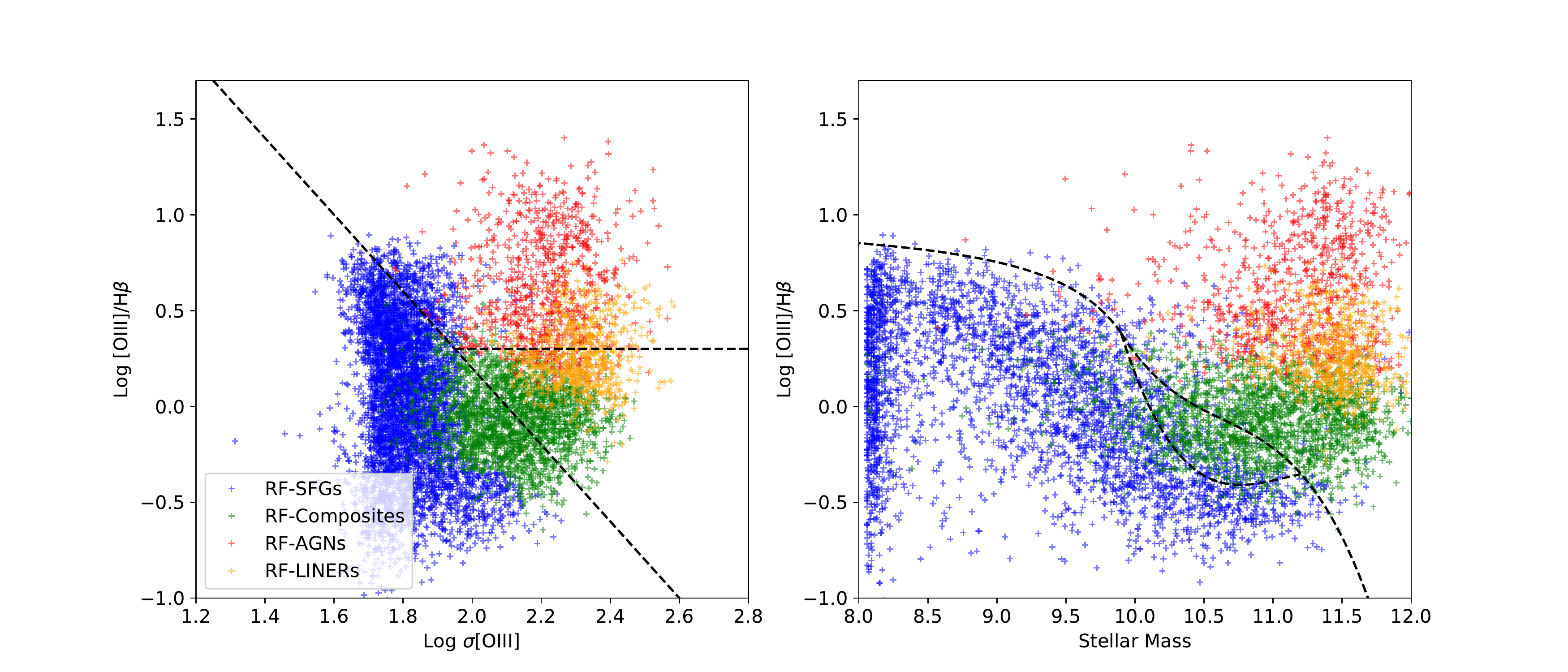}%{KEx-MEx.pdf}
\caption{ The RF-classified 0.32$<$z$<$0.8 galaxies on the kinematic--excitation (KEx; left panel; Zhang \& Hao 2018) and mass--excitation (MEx; right panel; Juneau et al. 2011) diagrams. The RF classification results are consistent with the demarcation lines proposed in those two works.  }
\label{KEx-MEx.fig}
\end{figure*}

In Figure~\ref{new-bpt.fig}, we show the RF-classified z$<$0.32 galaxies on the BPT diagram. The RF classifier reproduces the BPT diagram classification well, so Figures~\ref{bpt.fig} and \ref{new-bpt.fig} appear to be very similar.
In Figure~\ref{KEx-MEx.fig}, we plot the RF-classified 0.32$<$z$<$0.8 galaxies on the kinematic--excitation (KEx; Zhang \& Hao 2018) and mass--excitation (MEx; Juneau et al. 2011) diagrams. The stellar mass is drawn from the SDSS Galaxy Properties from the Wisconsin Group value-added catalog\footnote{\url{https://www.sdss.org/dr12/spectro/galaxy_wisconsin/}} (Chen et al. 2012). We chose the stellar mass derived using the Maraston et al. (2011) templates. The RF classification results are quite consistent with the KEx and MEx demarcation lines from Zhang \& Hao (2018) and Juneau et al. (2011), respectively. In terms of accuracy, the KEx diagram gives classification accuracies of 89\%, 75.6\%, and 81\% for SFGs, composites, and AGNs, respectively. The classification accuracies using the MEx diagram are 94.4\%, 47.8\%, and 54\% for SFGs, composites, and AGNs, respectively. By comparison, the accuracies of the RF classification for SFGs are 93.4\%, 69.4\%, 71.8\%. The KEx diagram achieves very good accuracy for composites and AGNs by sacrificing the accuracy of SFGs. To make an apples-to-apples comparison, we derive the ROC curves and AUC scores using the RF classifier but restrict it to only use the same 2 features as the KEx diagram: \oiii/\hb\ and $\sigma(\oiii)$. The AUC scores drop from 0.985, 0.966, 0.876, and 0.897 to 0.977, 0.936, 0.865, and 0.883 for SFGs, composites, AGNs and LINERs, respectively. The accuracies of the 2 feature RF classifier (see Table \ref{accuracy.tab}) are significantly lower than that of the 8 feature RF. Our results indicate that the RF classifier using 8 features gives as consistent of a classification as the BPT diagram, and it out-performs the KEx diagram and MEx diagram.

\section{Applying the Random Forest Classifier to 0.32$<$z$<$0.8 Emission Line Galaxies}

We apply the Random Forest classifier trained in Section~\ref{rf.sec} to 49,272 0.32$<$z$<$0.8 emission line galaxies in Section~\ref{highz_sample.sec}. We use the kcorrect package to k-correct the photometry to z=0.1 $u$, $g$, $r$, $i$, and $z$ magnitudes. \oiii/\hb, \oii/\hb, $\sigma_{\oiii}$, and $\sigma_*$ are measured using eBOSS spectra. The RF classifies 23,919 galaxies as star-forming, 13,536 as composites, 9,448 as AGNs, and 2,369 as LINERs.  The ideal method to test our classifications would be to observe the galaxies with near-IR spectra to cover the rest-frame optical wavelength range so that a BPT classification is possible
because this is the only way to accurately classify the sample into the four subtypes.  This is will possible when spectra from surveys like MOSFIRE Deep Evolution Field Survey (MOSDEF; Kriek et al. 2015; Sanders et al. 2016) become publicly available. In lieu of large numbers of near-IR spectra, we compare the stacked spectra of the four subtypes using the stacking code developed in Comparat et al. (2016)\footnote{\url{https://github.com/JohanComparat/pySU/blob/master/galaxy/python/SpectraStackingEBOSS.py}}. The stacked spectrum of each of the four subtypes should be significantly different from each other, and they should be generally consistent with their z$<$0.32 counterparts.

\begin{figure*}
\includegraphics[width=1\textwidth]{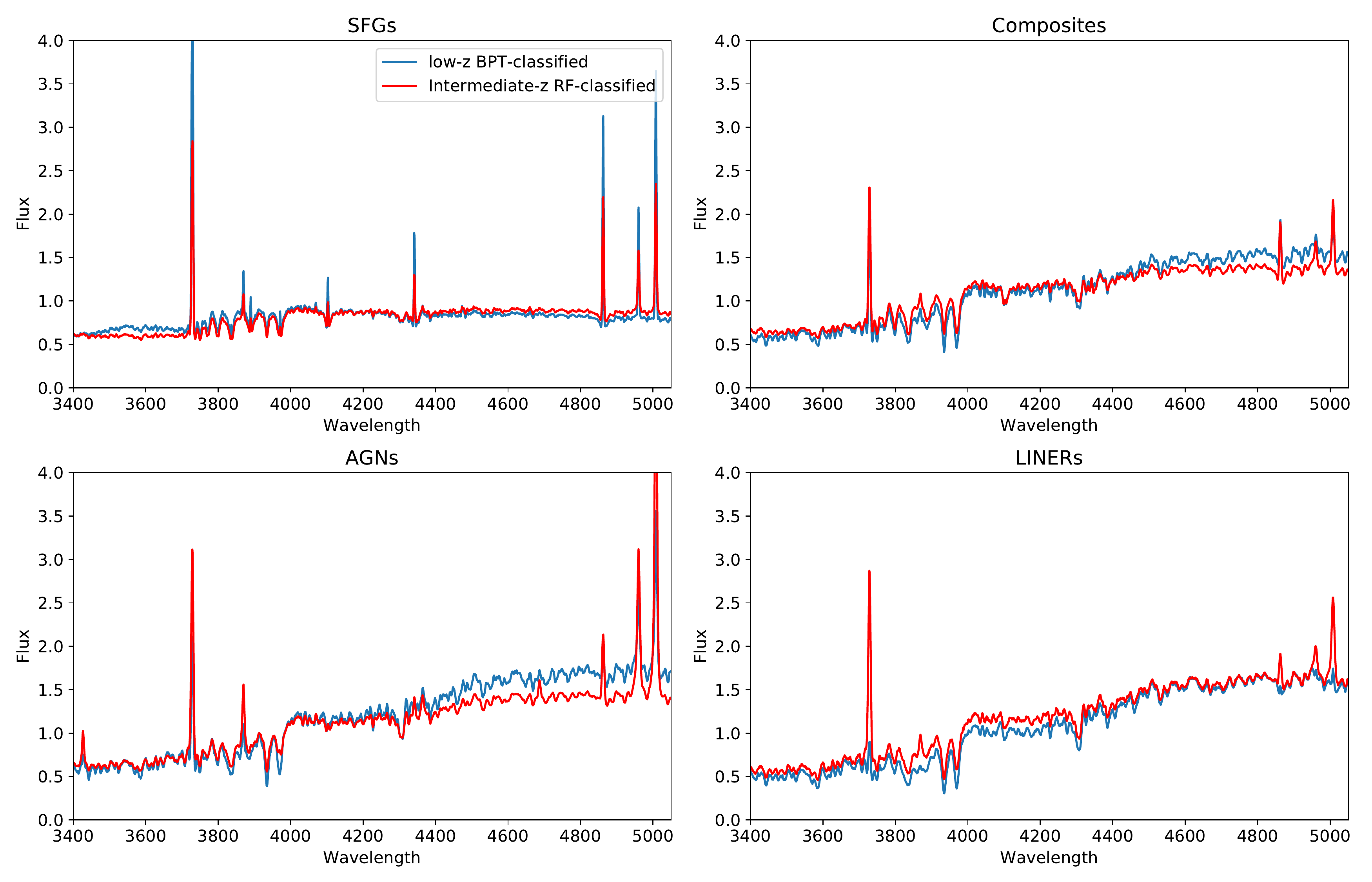} %stack_high_low.pdf
\caption{ A comparison of stacked spectra of 0.32$<$z$<$0.8 SFGs, composites, AGNs, and LINERs classified using the Random Forest classifier described in Section~\ref{rf.sec} shown in red and the BPT-classified z<0.32 SFGs, composites, AGNs, and LINERs shown in blue.}
\label{stack.fig}
\end{figure*}

Figure~\ref{stack.fig} shows the rest-frame 3400-5050\AA\ stacked spectra of RF-classified 0.32$<$z$<$0.8 SFGs, composites, AGNs, and LINERs and their z$<$0.32 counterparts classified using the BPT diagrams.
SFGs show prominent absorption features and the least steep continuum. AGNs show a steeper continuum than SFGs and prominent emission lines. Composites have features in between those of SFGs and AGNs, as expected. LINERs show the steepest spectrum and also significant emission lines. The spectral shape of RF-classified sources are highly consistent with BPT-classified low redshift ELGs of the same subtype. The \oiii/\hb\ ratios are consistent with expectations, too. This strongly suggests that the RF classifier is correctly classifying the four subtypes of ELGs.

Despite these consistencies, there are noticeable differences between the RF-classified and BPT-classified samples. The equivalent width of RF-classified composites and LINERs are significantly higher than their z$<$0.32 BPT-classified counterparts. There are at least two reasons for this difference.  First, the sample selection criteria require that the signal-to-noise ratios of \hb\ and \oiii\ to be greater than 3, effectively selecting stronger emission line galaxies at intermediate redshift than at low redshift. Second, the intermediate redshift composite and LINER groups are contaminated by AGNs and SFGs, whose much stronger emission lines will bias stacked spectra. This contamination does not dominate the spectra because the \oiii/\hb\ ratios are consistent with the low redshift values. Thus, we conclude that the selection effect is the main reason for the difference in the equivalent widths of the low and intermediate redshift composites and LINERs. 

%\section{Discussion}
%\label{discussion.sec}
%\subsection{Errors in Measurements}

\section{Conclusions}
In this paper, we consider the classification of intermediate redshift emission line galaxies using supervised machine learning classification algorithms. We use measurements available for optical spectra of galaxies at z$<$0.8: \oiii/\hb, \oii/\hb, \oiii\ line width ($\sigma(\oiii)$), stellar velocity dispersion ($\sigma_*$), u-g, g-r, r-i, and i-z color as input. A z$<0.3$ emission line galaxy sample classified and labeled using standard optical diagnostic diagrams is selected as training sample. We use $k$-nearest neighbors (KNN), support vector classifier (SVC), random forest (RF), and a multi-Layer perceptron neural network (MLP-NN) to train models that predict which class a galaxy belongs to given a set of input. Receiver operating characteristic (ROC) curve and area under curve (AUC) score are used to quantify the distinguishing power of the different classifiers. RF has the best AUC score for classifications of all four subtypes, while the relative ranking of the other three algorithms in both AUC scores and accuracies is MLP$>$SVC$>$KNN. The RF classification accuracies are 93.4\%, 69.4\%, 71.8\%, and 65.7\% for star-forming galaxies, composites, AGNs, and LINERs, respectively. The three most important features are \oiii/\hb, $\sigma_{\oiii}$, and g-r. Reducing the input to the four spectroscopic features results in slightly degraded accuracies of 92.3\%, 63.7\%, 67.3\%, and 60.8\%. The stacked spectra of the four subtypes classified using the RF model are consistent with the stacked spectra of low redshift BPT-classified ELGs. The machine learning classification tool will play an important role in emission line galaxy physics in upcoming large sky surveys like DESI, PFS, and 4MOST.

\section*{Acknowledgements}
KZ thanks Xiaosheng Huang for helpful discussion on machine learning technics. Funding for the Sloan Digital Sky Survey IV has been provided by the Alfred P. Sloan Foundation, the U.S. Department of Energy Office of Science, and the Participating Institutions. SDSS- IV acknowledges support and resources from the Center for High-Performance Computing at the University of Utah. The SDSS web site is \url{www.sdss.org}.

SDSS-IV is managed by the Astrophysical Research Consortium for the Participating Institutions of the SDSS Collaboration including the Brazilian Participation Group, the Carnegie Institution for Science, Carnegie Mellon University, the Chilean Participation Group, the French Participation Group, Harvard-Smithsonian Center for Astrophysics, Instituto de Astrof\'isica de Canarias, The Johns Hopkins University, Kavli Institute for the Physics and Mathematics of the Universe (IPMU) / University of Tokyo, Lawrence Berkeley National Laboratory, Leibniz Institut f\"ur Astrophysik Potsdam (AIP), Max-Planck-Institut f\"ur Astronomie (MPIA Heidelberg), Max-Planck-Institut f\"ur Astrophysik (MPA Garching), Max-Planck-Institut f\"ur Extraterrestrische Physik (MPE), National Astronomical Observatory of China, New Mexico State University, New York University, University of Notre Dame, Observat\'ario Nacional / MCTI, The Ohio State University, Pennsylvania State University, Shanghai Astronomical Observatory, United Kingdom Participation Group, Universidad Nacional Aut\'onoma de M\'exico, University of Arizona, University of Colorado Boulder, University of Oxford, University of Portsmouth, University of Utah, University of Virginia, University of Washington, University of Wisconsin, Vanderbilt University, and Yale University.
\newline

%%%%%%%%%%%%%%%%% Table 1 %%%%%%%%%%%%%%%%%%%%%%%%%%%%%%%%%%%%%%%

\begin{table*}[]
%\centering
\topmargin 0.0cm
\evensidemargin = 0mm
\oddsidemargin = 0mm
\scriptsize %\small %\tiny %
\caption{The AUC scores for different machine learning classifiers}
\label{auc.tab}
\medskip
\vfill
\begin{tabular}{l|c c c c | c c c }
\hline \hline
Type             &  Star-Forming Galaxies  & Composites &  AGNs & LINERs & Average\\
                   &                                          &                    &                   &  \\
(1)              & (2)                                     & (3)               & (4)            & (5) & (6)       \\
\hline \hline  %%
KNN        & 0.964$\pm$0.004      & 0.878$\pm$0.010       & 0.860$\pm$0.007           & 0.865$\pm$0.008   & 0.892 \\
SVC       & 0.968$\pm$0.005      & 0.881$\pm$0.014       & 0.861$\pm$0.009           & 0.869$\pm$0.009   & 0.895 \\
MLP       & 0.973$\pm$0.003      & 0.896$\pm$0.009       & 0.880$\pm$0.004           & 0.877$\pm$0.009   & 0.906 \\
RF         & 0.985$\pm$0.001      & 0.966$\pm$0.004       & 0.876$\pm$0.008           & 0.897$\pm$0.004   & 0.931\\
RF (4 features) & 0.981$\pm$0.002      & 0.952$\pm$0.003       & 0.870$\pm$0.009           & 0.890$\pm$0.007   & 0.923\\
RF (2 features) & 0.977$\pm$0.003      & 0.936$\pm$0.010       & 0.865$\pm$0.007           & 0.883$\pm$0.006   & 0.915\\
 \hline
\end{tabular}
\medskip
\vfill
{\normalsize ~Columns: (1) Classifier type.
 (2)Star-forming galaxies AUC score. (3) Composite galaxies classification AUC score.
(4) LINERs classification AUC score. (5) AGNs classification AUC score.  (6) Average AUC score.}\\
 %\end{sidewaystable*}
\end{table*}

%%%%%%%%%%%%%%%%% Table 2 %%%%%%%%%%%%%%%%%%%%%%%%%%%%%%%%%%%%%%%

\begin{table*}[]
%\centering
\topmargin 0.0cm
\evensidemargin = 0mm
\oddsidemargin = 0mm
\scriptsize %\small %\tiny %
\caption{Accuracies for different machine learning classifiers}
\label{accuracy.tab}
\medskip
\vfill
\begin{tabular}{l|c c c c | c c c }
\hline \hline
Type             &  Star-Forming Galaxies  & Composites &  AGNs & LINERs  & Overall\\
                   &                                          &                    &                     & \\
(1)              & (2)                                     & (3)               & (4)                 & (5) & (6)   \\
\hline \hline  %%
KNN                 & 0.934$\pm$0.001 & 0.596$\pm$0.006 & 0.760$\pm$0.008 & 0.516$\pm$0.009  & 0.702\\
SVC                 & 0.926$\pm$0.003 & 0.638$\pm$0.008 & 0.794$\pm$0.004 & 0.608$\pm$0.007   & 0.741 \\
MLP                  & 0.937$\pm$0.004 & 0.688$\pm$0.017 & 0.768$\pm$0.022 & 0.611$\pm$0.019  & 0.750 \\
RF                    & 0.934$\pm$0.001 & 0.694$\pm$0.005 & 0.718$\pm$0.011 & 0.657$\pm$0.010  & 0.751\\
RF (4 features) &0.923$\pm$0.002 & 0.637$\pm$0.008  & 0.673$\pm$0.005  & 0.608$\pm$0.009  & 0.710 \\
RF (2 features) &0.909$\pm$0.002 & 0.594$\pm$0.007  & 0.564$\pm$0.013  & 0.482$\pm$0.013  & 0.637\\
 \hline
\end{tabular}
\medskip
\vfill
{\normalsize ~Columns: (1) Classifier type.
 (2)Star-forming galaxies classification accuracy. (3) Composite galaxies classification accuracy.
(4) LINERs classification accuracy. (5) AGNs classification accuracy.  (6) Overall classification accuracy.}\\
 %\end{sidewaystable*}
\end{table*}

%%%%%%%%%%%%%%%%% Table 3 %%%%%%%%%%%%%%%%%%%%%%%%%%%%%%%%%%%%%%%

\begin{table*}[]
%\centering
\topmargin 0.0cm
\evensidemargin = 0mm
\oddsidemargin = 0mm
\scriptsize %\small %\tiny %
\caption{Confusion Matrix for the Random Forest Classifier}
\label{confusion.tab}
\medskip
\vfill
\begin{tabular}{l|c c c c | c c c }
\hline \hline
Type             &  Star-Forming Galaxies  & Composites &  AGNs & LINERs  \\
                   &                                          &                    &                     & \\
(1)              & (2)                                     & (3)               & (4)                 & (5)   \\
\hline \hline  %%
 True SFGs       & 0.937        & 0.060     & 0.003       & 0.000      \\
 True Composites                        & 0.238        & 0.683      & 0.018     & 0.061      \\
True AGNs                                  & 0.018         & 0.089     & 0.705      & 0.188     \\
True LINERs                               & 0.004        & 0.284      & 0.067      & 0.644       \\
 \hline
\end{tabular}
\medskip
\vfill
{\normalsize ~Confusion matrix of the random forest classifier. Each row gives the probabilities that galaxies of a given subtype of ELG are classified as each of the four subtypes. Star-forming
galaxies are unlikely to confuse with the other subtypes. Composites have a 23.8\% probability to be confused with star-forming galaxies. AGNs have 18.8\%,  8.9\%, and 1.8\% probabilities to be classified as LINERs, composites and star-forming galaxies. LINERs are most likely to be misclassified as composites with 28.4\% probability, with 6.7\% and 0.4\% probabilities to be misclassified as AGNs and star-forming galaxies. }\\
 %\end{sidewaystable*}
\end{table*}

%Tianqi Chen and Carlos Guestrin. XGBoost: A Scalable Tree Boosting System. In 22nd SIGKDD Conference on Knowledge Discovery and Data Mining, 2016
%Olga Russakovsky*, Jia Deng*, Hao Su, Jonathan Krause, Sanjeev Satheesh, Sean Ma, Zhiheng Huang, Andrej Karpathy, Aditya Khosla, Michael Bernstein, Alexander C. Berg and Li Fei-Fei. (* = equal contribution) ImageNet Large Scale Visual Recognition Challenge. International Journal of Computer Vision, 2015.
%%%%%%%%%%%%%%%% end of Table 2 %%%%%%%%%%%%%%%
%%%%%%%%%%%%%%%%%%%%%%%%%%%%%%%%%%%%%%%%% figures

%\end{CJK}
\end{document}